\documentclass{ragtime} 

\title[Polarization modelling of the AGN unified scheme]
      {Modelling the polarization dichotomy\\
       of Active Galactic Nuclei}

\author[Ren\'e W. Goosmann]
       {Ren\'e W. Goosmann\\
        Astronomical Institute, Academy of Sciences\\
        Bo{\v c}n\'i II 1401a, CZ--14131 Prague, Czech Republic\\
        \Email{goosmann@astro.cas.cz}} 

\begin{document}

\begin{abstract}
  I present polarization modelling of Active Galactic Nuclei in the optical/UV
  range. The modelling is conducted using the Monte-Carlo radiative transfer
  code {\sc Stokes}, which self-consistently models the polarization signature
  of a complex model arrangement for an active nucleus. In this work I
  include three different scattering regions around the central source: an
  equatorial electron scattering disk, an equatorial obscuring dusty torus,
  and polar electron scattering cones. I investigate the resulting
  dependencies of the V-band polarization for different optical depths of
  the scattering cones, different dust compositions inside the torus, and
  various half-opening angles of the torus/polar cones. The observed
  polarization dichotomy can be successfully reproduced by the model.
\end{abstract}


\section{Introduction}
\label{intro}

The research of Active Galactic Nuclei (AGN) started with the discovery of
separate types of objects that nowadays we gather under the AGN-class: Seyfert
galaxies, radio galaxies, quasars, blazars, etc... An important common
property among the various types is the very strong luminosity produced inside
a small spatial region at the centre of the host galaxy. The standard model of
the AGN phenomenon assumes accretion onto a supermassive black hole as the
fundamental mechanism for producing the strong radiation. The central black
hole and the accretion flow are surrounded by several additional media such as
the broad line and narrow line regions. A key assumption of this so-called
{\it unified scheme} for AGN is the existence of a dusty torus \citep[see
e.g.][]{antonucci1993}. It divides AGN into two classes: ``type-1'' objects,
which are seen close to face-on, and ``type-2'' objects seen rather
edge-on. In type-1 AGN the central energy source and the broad line region can
be seen directly, whilst in type-2 AGN the torus blocks the view toward the
centre.

Current observational technology in the optical and UV waveband does not allow
to resolve the inner regions of AGN. However, the light of AGN is polarized
over a broad wavelength range, which allows to put important constraints on
the geometry of the emitting and scattering regions. When light is scattered,
the angle of polarization depends on the direction of the last scattering, so
one expects the angle of polarization to be related to the structure of the
AGN. \citet{stockman1979} made the seminal discovery that for
low-polarization, high optical luminosity, radio-loud AGN, the optical
polarization position angles tend to align {\it parallel} to the large-scale
radio structure. \citet{antonucci1982} pointed out that whilst many radio
galaxies showed a similar parallel alignment of the polarization and radio
axes, there was, unexpectedly, a population showing a {\it perpendicular}
relationship. It was subsequently shown \citep{antonucci1983} that
relatively-radio-quiet Seyfert galaxies show a similar dichotomy between the
predominantly, but not exclusively, parallel polarization in face-on type-1
Seyferts and the perpendicular polarization of type-2 Seyferts
\citep[see][~for reviews]{antonucci1993,antonucci2002}.

Applying the radiative transfer code {\sc Stokes}, we have presented
theoretical modelling of individual scattering regions in AGN
\citep{goosmann2007a, goosmann2007b, goosmann2007c}. In these papers, we
considered dusty torii, polar electron cones, and equatorial scattering wedges
individually calculating their polarization signatures for various viewing
angles. For the present proceedings note I expand on this type of modelling
combining the individual regions to obtain an approach to the unified scheme
of AGN. With {\sc Stokes} such modelling is done consistently as the code
automatically includes the effects of multiple scattering. In
Sect.~\ref{sec:code}, I briefly summarize the basic properties of {\sc
  Stokes}. In Sect.~\ref{sec:model}, I describe the model setup. The modelling
results are then presented in Sect.~\ref{sec:results} and discussed in
Sect.~\ref{sec:discuss}.

\section{The radiative transfer code Stokes}
\label{sec:code}

\begin{figure}
  \centering
  \includegraphics[width=0.7\linewidth]{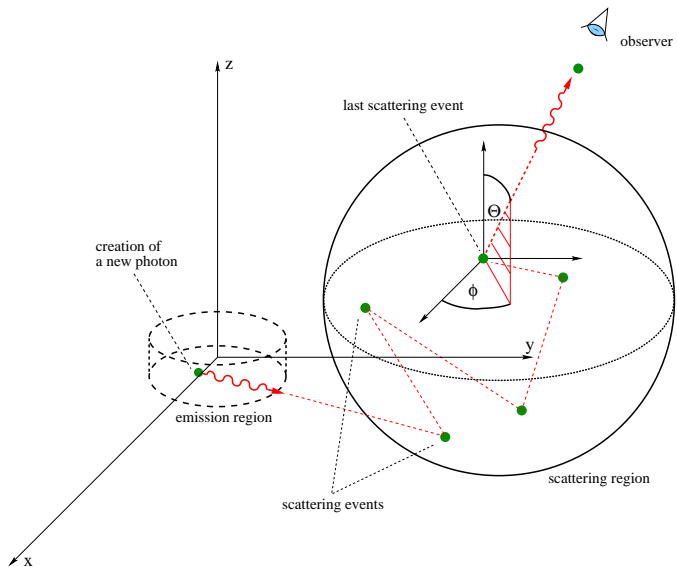}
  \caption{A photon working its way through the model space.}
  \label{fig:model-space}
\end{figure}

The computer program {\sc Stokes} performs simulations of radiative transfer,
including the treatment of polarization, for AGN and related objects. The code
is publicly available and 100\%
shareware\footnote{http://www.stokes-program.info/}. It is based on the Monte
Carlo method and follows single photons inside the source region through
various scattering processes until they become absorbed or escape from the
model region (Fig.~\ref{fig:model-space}). 

Photons are created inside the source regions, which can be defined by
different geometries. The continuum radiation is set by the index $\alpha$ of
an $F_\nu \propto \nu^{-\alpha}$ power law. The Stokes vectors of the emitted
photons are initially set to the values of completely unpolarized
light. Various scattering regions can be arranged around the source
regions. The program offers e.g. toroidal, cylindrical, spherical or conical
shapes. These regions can contain free electrons or dust consisting of
``astronomical silicate'' and graphite. A photon works its way through the
model region and generally undergoes several scattering events. The emission
directions, path lengths between scattering events, and the scattering angles
are sampled by Monte Carlo routines based on classical intensity
distributions. During each scattering event the Stokes vector is changed by
multiplication with the corresponding Mueller matrix. For dust scattering,
absorption is important, and a large fraction of the photons does not reach
the virtual observer. The relevant cross sections and matrix elements for dust
scattering and absorption are computed on the basis of Mie theory applied to
size distributions of spherical graphite and silicate grains.

When a photon escapes from the model region, it is registered by a web of
virtual detectors arranged in a spherical geometry around the source. The flux
and polarization information of each detector is obtained by adding up the
Stokes parameters of all detected photons. If the model is completely axially
symmetric these can be azimuthally integrated and, if there is plane symmetry,
the top and bottom halves are combined. The object can be analyzed in total
flux, in polarized flux, percentage of polarization, and the position angle at
each viewing angle.

\section{Modelling the unified scheme}
\label{sec:model}

\begin{figure}
  \begin{center}
    \includegraphics[width=0.5\linewidth]{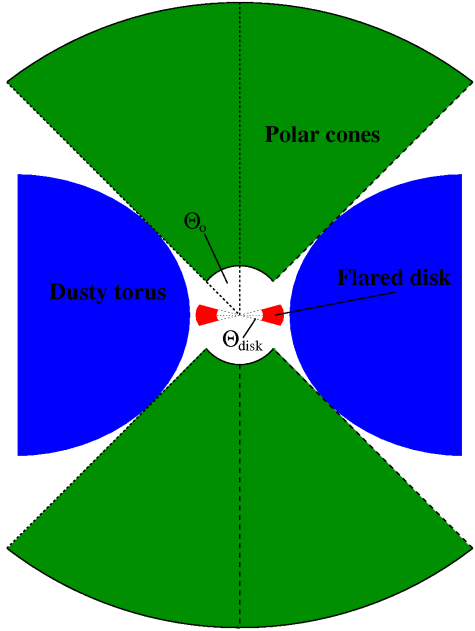}
  \end{center}
  \caption{Illustration of the setup for a unified scheme model of AGN. The
    central source is surrounded by a flared electron scattering disk (wedge),
    a dusty torus, and by polar electron cones. \label{fig:geom}}
\end{figure}

Our setup for the united model of AGN is shown in Fig.~\ref{fig:geom}. We
include the equatorial dusty torus and the polar electron scattering cones. In
addition to that an equatorial electron scattering wedge is defined. Such a
region produces the correct (parallel) polarization of type-1 AGN. The
polarization properties of flat equatorial scattering disks have been
investigated in a series of papers by \citet{young2000} and
\citet{smith2002,smith2004,smith2005} as well as in \citet{goosmann2007a}.

We assume that the central source of the AGN is point-like and emits a flat
intrinsic spectrum around $\lambda = 5500$~\AA. We define a half-opening
angle, $\theta_{\rm disk}$, of the flared electron disk of $\theta_{\rm disk} =
25^\circ$. For this half-opening angle a high percentage of type-1
polarization is expected \citep{goosmann2007a}. The radial Thomson optical
depth of the wedge is set to unity. The half-opening angle $\theta_0$ of the
torus and the cone are set equal, which corresponds to the interpretation that
the ionized outflow is collimated by the torus. We consider two cases for the
Thomson optical depth of the scattering cones, which is measured along the
symmetry axis of a single cone and set to $\tau_{\rm cone} = 0.01$ and
$\tau_{\rm cone} = 0.1$ respectively. The radial optical depth of the dusty
torus in the equatorial plane is set to 750 for the V-band. The dust models
(table~\ref{tab}) assume a mixture of graphite and ``astronomical silicate''
and a grain radii distribution $n(a) \propto a^\alpha_s$ between $a_{\rm min}$
and  $a_{\rm max}$.

\begin{table}
  \begin{center}
    \caption{Parameterization of the dust models}
    {\footnotesize
    \label{tab}
    \begin{tabular}{lccccc}\hline
      Type & Graphite & Silicate & $a_{\rm min}$ & $a_{\rm max}$ &
      $\alpha_s$\\\hline
      Galactic & 62.5\% & 37.5\% & 0.005$\, \mu{\rm m}$ & 0.250$\, \mu{\rm m}$
               & $-3.5$\\
      AGN      &  85\%  &  15\%  & 0.005$\, \mu{\rm m}$ & 0.200$\, \mu{\rm m}$
               & $-2.05$\\\hline
    \end{tabular}
    }
  \end{center}
\end{table}

The ``Galactic dust'' model reproduces the interstellar extinction for
$R_{\rm V} = 3.1$ whilst the ``AGN dust'' parameterization is obtained from
quasar extinction curves derived by \citet{gaskell2004}. This latter dust type
favours larger grain sizes. Using {\sc Stokes} we consistently model the
resulting polarization spectrum of the entire model setup for various
inclinations, $i$, of the observer and for four different values of $\theta_0$
between $30^\circ$ and $45^\circ$.

\section{Results}
\label{sec:results}

\begin{figure}[t]
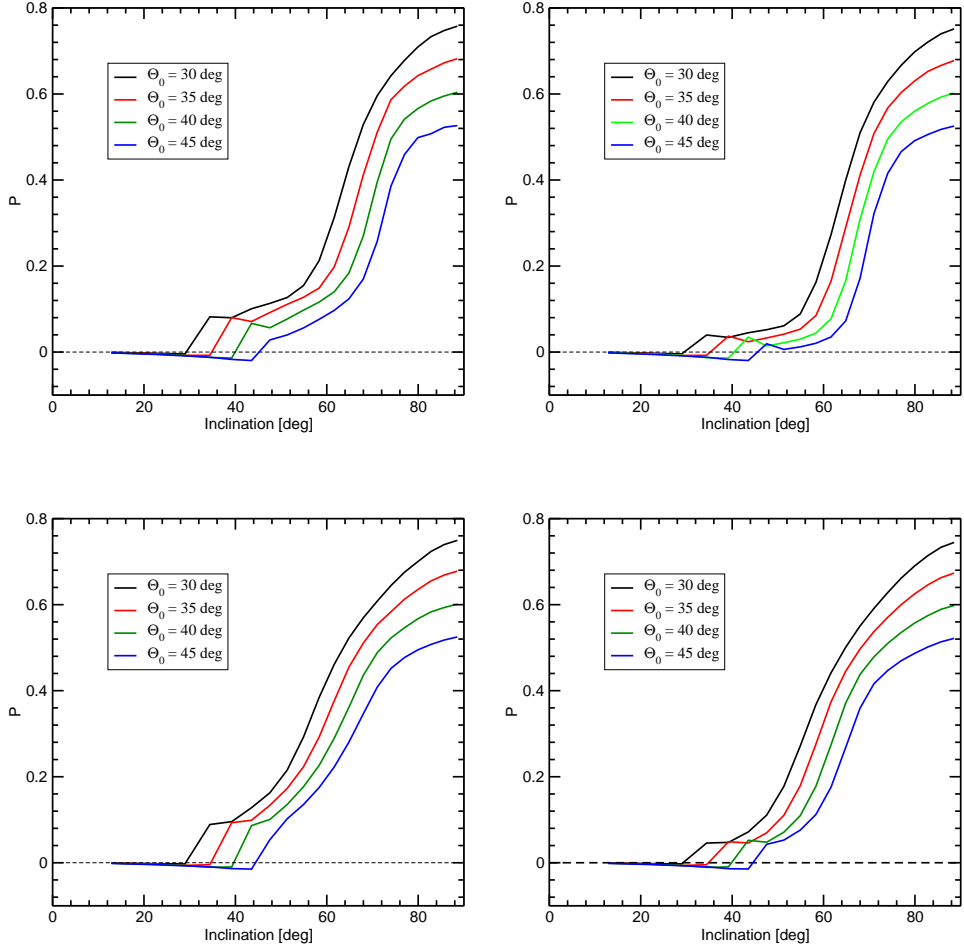

  \begin{center}
    \vskip 0.6cm
    \includegraphics[width=0.48\linewidth]{MilkyWay-AGN-largescale.eps}
    \hfill
    \includegraphics[width=0.48\linewidth]{AGN-dust-AGN-largescale.eps}
    \vskip 1cm
    \includegraphics[width=0.48\linewidth]{MilkyWay-AGN-largescale-tau0.1.eps}
    \hfill
    \includegraphics[width=0.48\linewidth]{AGN-dust-AGN-largescale-tau0.1.eps}
    \vskip 0.6cm
  \end{center}
  \caption{Polarization degree $P$ at $5500 \, \rm{\AA}$ as a function of the
    disk inclination $i$. The upper panels denote an optical depth $\tau =
    0.01$ of the polar electron cones, the lower ones denote $\tau = 0.1$. On
    the left side, the results for a torus with Galactic dust are given, on
    the right side the results are for AGN dust. From left to write the four
    curves of each panel mark increasing half-opening angles $\theta_0$ of the
    system. Positive values of the polarization degree denote an orientation
    of the polarization position angle, which is perpendicular to the symmetry
    axis, negative values stand for parallel polarization. \label{fig3}}
\end{figure}

\begin{figure}[t]
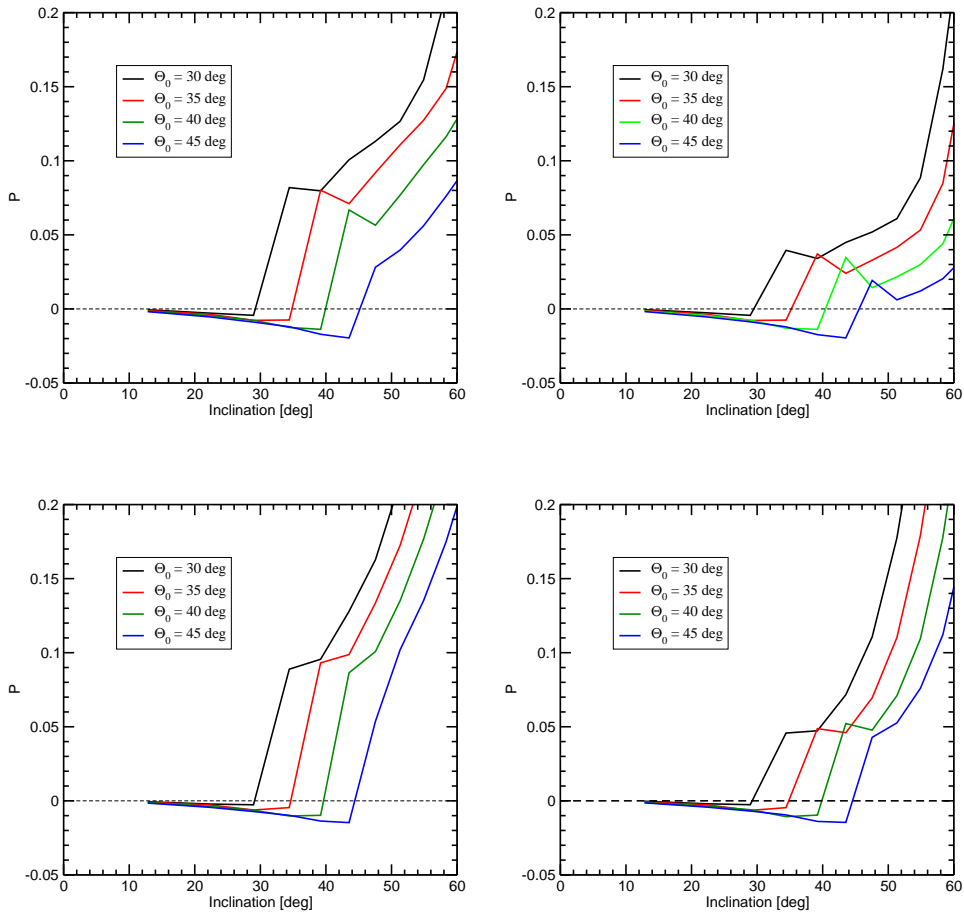

  \begin{center}
    \vskip 0.6cm
    \includegraphics[width=0.48\linewidth]{MilkyWay-AGN-smallscale.eps}
    \hfill
    \includegraphics[width=0.48\linewidth]{AGN-dust-AGN-smallscale.eps}
    \vskip 1cm
    \includegraphics[width=0.48\linewidth]{MilkyWay-AGN-smallscale-tau0.1.eps}
    \hfill
    \includegraphics[width=0.48\linewidth]{AGN-dust-AGN-smallscale-tau0.1.eps}
    \vskip 0.6cm
  \end{center}
  \caption{Same as in Fig.~\ref{fig3} but zoomed in and limited to disk
    inclinations of $60^\circ$. \label{fig4}} 
\end{figure}

We investigate the dependence of the polarization in the visual band on the
half-opening angle of the dusty torus/polar cones. In Fig.~\ref{fig3} and
Fig.~\ref{fig4} we show the resulting percentage of polarization, $P$, versus
$i$ for the two values of $\tau$ and for the two types of dust. The relation
has a similar shape for all cases shown and reproduces the observed
type-1/type-2 polarization dichotomy: the polarization position angle is
oriented parallel to the projected symmetry axis when the line of sight is
above the horizon of the torus, i.e. for $i < \theta_0$, and switches to a
perpendicular orientation for $i > \theta_0$. In the figures, the two different
orientations of the polarization vector are denoted by negative (type-1) and
positive (type-2) values of $P$. The type-1 values of $P$ are moderate and
reach maximum absolute values of $\sim 2.5\%$. They rise with $i$ until the
polarization vector switches to the type-2 orientation. In the type-2 case $P$
continues to increase with $i$ and saturates for edge-on viewing angles at a
level that depends on $\theta_0$.

The combined effect of all scattering regions on the total polarization value
can be partly understood from the results we obtained when modelling the
individual regions in \citet{goosmann2007a}. However, the fact that all
regions are radiatively coupled adds more complexity to the model. The polar
scattering regions have a strong impact on the result, in particular for
type-2 viewing angles. With increasing $\theta_0$ the resulting type-2
polarization becomes lower because it is averaged over a broader distribution
of polarization position angles. An increasing optical depth of the cones
raises $P$ for the type-2 case because more photons are scattered by the
cones. 

For nearly face-on viewing directions, the polar cones have less impact as
they cause mainly forward or backward scattering producing low
polarization. In these cases the resulting polarization is mainly determined
by the geometry and optical depth of the equatorial scattering disk. However,
these two regions compete against each other, as they produce different
orientations of the polarization vector. For higher optical depth the impact
of the polar cones becomes stronger and lowers the resulting type-1
polarization, as can be seen when comparing the top with the bottom panels in
Figs.~\ref{fig3}~and~\ref{fig4}.

In the central parts of the model region, the optically thick torus, and the
scattering wedge are strongly interconnected by multi-scattering. This
explains the significant impact of the dust composition and grain size
distribution on the resulting polarization profile. For AGN dust the obtained
type-2 polarization percentages for intermediate viewing angles are lower than
for the Galactic dust torus. For this range of $i$ the reflection off the
torus has an important influence while toward edge-on values of $i$ the polar
scattering is again more important.

\section{Summary and discussion}
\label{sec:discuss}

In this proceedings note, we have investigated the optical polarization
imprint of an active nucleus. Our model is geared toward the unified scheme of
AGN including equatorial and polar electron scattering regions and an
obscuring dusty torus. Evaluating the polarization percentage and position
angle for various disk inclinations we succeed to reproduce the observed
polarization dichotomy between type-1 and type-2 AGN. However, the presence of
the dichotomy is sensitive to the model parameters, as it is the result of the
competing type-1 polarization produced by the equatorial wedge on the one
hand, and the type-2 polarization caused by the torus and the polar cones on
the other.

\citet{smith2004} pointed out that the competition between the equatorial and
polar scattering explains the special population of Seyfert-1 galaxies that
shows type-2 polarization. They are considered to be dominated by polar
instead of equatorial scattering. In our model, we have set the half-opening
angle and the optical depth of the equatorial wedge in such a way that a
maximum type-1 polarization percentage is obtained. We then vary the optical
depth of the polar scattering cones. The resulting distribution of the
polarization position angle as a function of the inclination must correspond
to the observed number densities of Seyfert-1 galaxies that are dominated by
polar and by equatorial scattering. In principle, it is thus possible to put
constraints on the optical depth of the polar cones.

But our modelling shows that there is also a significant impact of the dusty
torus on the resulting polarization, especially for intermediate viewing
angles. The resulting polarization changes with the dust type. From the given
number density of AGN with different spectral and polarization types it is
thus not straightforward to find relations between the properties of the
various scattering regions. It rather requires more detailed modelling over a
broader spectral range and within a larger parameter space than presented
here. We intend to conduct such investigations in the future.

\ack This work was supported by the Centre for Theoretical Astrophysics
through research grant LC06014.

\bibliography{\jobname}

\end{document}